\documentclass[twocolappendix, apj]{emulateapj}
\usepackage{natbib}
\usepackage{epsfig}
\usepackage{amsmath}

\def\hMpc{~h^{-1}{\rm Mpc}}

\slugcomment{Submitted to ..; Accepted for publication ..}

\shorttitle{Detection of BAO Using Galaxy Clusters}
\shortauthors{HONG ET AL.}

\begin{document}

\title{The correlation function of galaxy clusters and detection of baryon acoustic oscillations}

\author{
T.\ Hong, 
J. L.\ Han, 
Z. L.\ Wen, 
L.\ Sun, 
and
H.\ Zhan 
}
\affil{
National Astronomical Observatories, Chinese Academy
  of Sciences, 20A Datun Road, Chaoyang District, Beijing 100012,
  China. hjl@nao.cas.cn}

\begin{abstract}
We calculate the correlation function of 13\,904 galaxy clusters of $z \leq 0.4$ 
selected from the cluster catalog of Wen,
Han \& Liu.  The correlation function can be fitted with a power-law
model $\xi(r)=\left(r/R_{0}\right)^{-\gamma}$ on the scales of
$10\hMpc \leq r \leq 50\hMpc$, with a larger correlation length of
$R_0=18.84\pm0.27 \hMpc$ for clusters with a richness of $R\geq15$ and
a smaller length of $R_0=16.15\pm0.13 \hMpc$ for clusters with a
richness of $R \geq 5$. The power law index of $\gamma=2.1$ is
found to be almost the same for all cluster subsamples.  A
pronounced baryon acoustic oscillations (BAO) peak is detected at
$ r \sim 110 \hMpc $ with a significance of $\sim 1.9\sigma$. By
analyzing the correlation function in the range of $20\hMpc \leq r
\leq 200 \hMpc$, we find the
constraints on distance parameters are $D_v(0.276)=1077\pm55(1\sigma)
{~\rm Mpc}$ and $h=0.73\pm0.039(1\sigma)$, which are consistent with
the WMAP 7-year cosmology. However, the BAO signal from the cluster sample 
is stronger than expected and leads to a rather low matter density 
$\Omega_m h^2=0.093\pm0.0077(1\sigma)$, which deviates from 
the WMAP 7-year result by more than $3\sigma$. The correlation function of 
the GMBCG
cluster sample is also calculated and our detection of the BAO feature
is confirmed.
\end{abstract}


\section{Introduction}
One of the most important tasks of modern redshift surveys
\citep[e.g.][]{yaa+00,cdm+01} is to reveal the large-scale structure
of the universe.  Observations have shown that galaxies are not
uniformly distributed in the universe \citep{hmc+03,zwz+04}. They not
only cluster on the scales of Mpc \citep[e.g.][]{aco89,whl09} but also
are embedded in large-scale diverse structures which have scales of
tens of Mpc, exhibited as filaments, walls and voids
\citep[][]{gh89,gjs+05}. Large-scale baryon fluctuations bear the
imprint of acoustic oscillations in the tightly coupled baryon-photon
fluid prior to the epoch of recombination \citep{py70,sz70}.

To reveal the large-scale baryon acoustic oscillation (BAO) feature in
the universe \citep{eis02, eis05, mar09}, a large sample of tracers in
a huge volume of the order $1~ h^{-3}{\rm Gpc^{3}}$ need to be
observed.  Galaxies, especially the Luminous Red Galaxies (LRGs), are
the most popular tracers used. The Sloan Digital Sky Survey
\citep[SDSS, ][]{yaa+00} has observed the spectra for an unprecedently
large sample of galaxies. Using the LRG sample from the SDSS Data
Release~3 (DR3), \citet{ezh+05} firstly detected the BAO signal ($
\sim 3.4 \sigma $) at a scale of $ r \sim100\hMpc$ in the correlation
functions of LRGs, which has recently been updated by \citet{pre+10}
and \citet{kbs+10} using new LRG samples from the SDSS DR7.
\citet{bdp+11} and \citet{bbc+11} detected the BAO features
  using the data from the WiggleZ and 6dF galaxy surveys,
  respectively.

Being the largest gravitationally bound systems in the universe,
galaxy clusters are formed from more massive halos than galaxies.
They are more strongly correlated in space than galaxies.  Clusters
can be identified from photometric survey data by visual inspection
\citep{abe58, aco89} or automatic cluster-finding algorithms if the
redshifts of galaxies are available \citep{gdl+03, ldg+04}. Because of
lack of spectroscopic data of galaxies, galaxy clusters previously
identified in limited volumes of the universe are hard to be used for
the BAO detection. \citet{mnb01} showed the power spectrum with a
possible BAO feature from the redshift data of the Abell/ACO clusters,
Infrared Astronomical Satellite (IRAS) point sources and the Automated
Plate Measurement (APM) clusters. Using the large sample of galaxy
redshift data obtained from spectral observations or estimated from
photometric data of SDSS \citep{yaa+00}, a large number of galaxy
clusters have been identified \citep{kma+07a, kma+07b, whl09} in the
vast volume up to redshift $z\sim0.6$. They can be used to trace the
large scale structure and constrain cosmological parameters, e.g. the
mass fluctuation on the scale of $8\hMpc$, $\sigma_8$ \citep{whl10}.
\citet{esf09} detected the BAO signature ($1.4\sigma \sim
1.7\sigma$) from the cluster correlation function using the maxBCG
cluster sample \citep{kma+07a, kma+07b}. \citet{hue10} also found the
BAO feature ($\sim 2.2 \sigma$) in the redshift-space power spectrum
of the maxBCG cluster sample.

\begin{figure*}[t]
\centering
\includegraphics[width=0.32\textwidth]{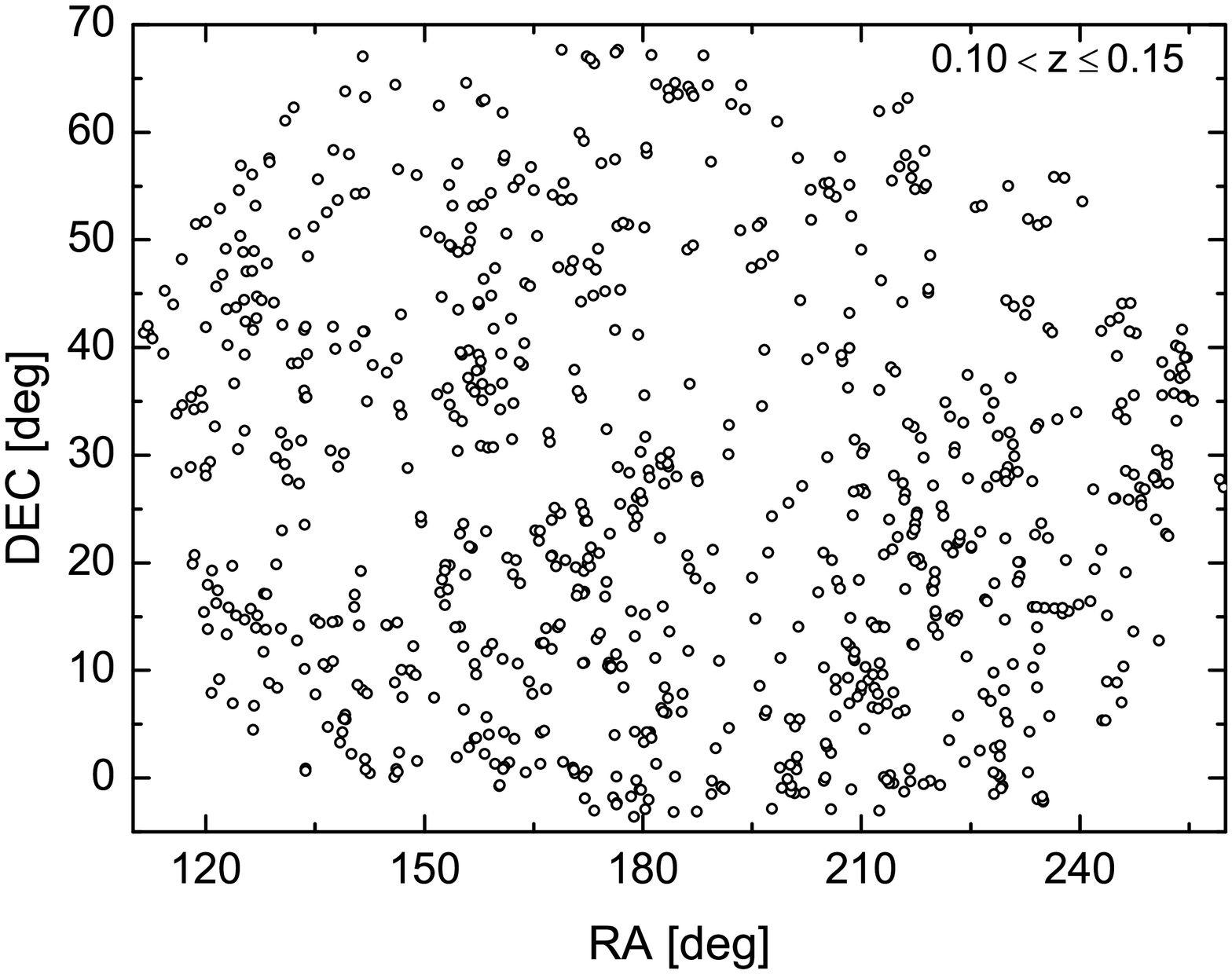}
\includegraphics[width=0.32\textwidth]{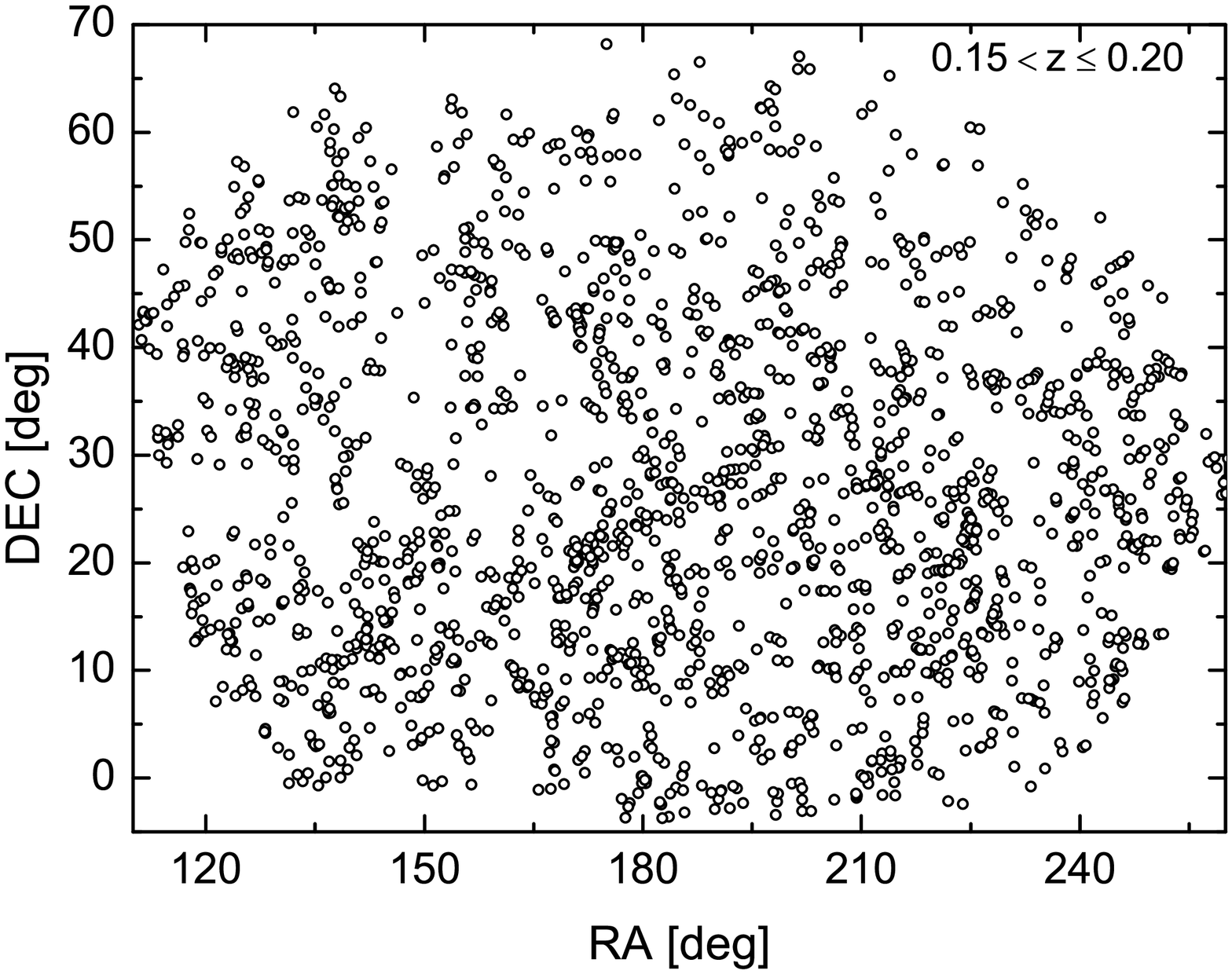}
\includegraphics[width=0.32\textwidth]{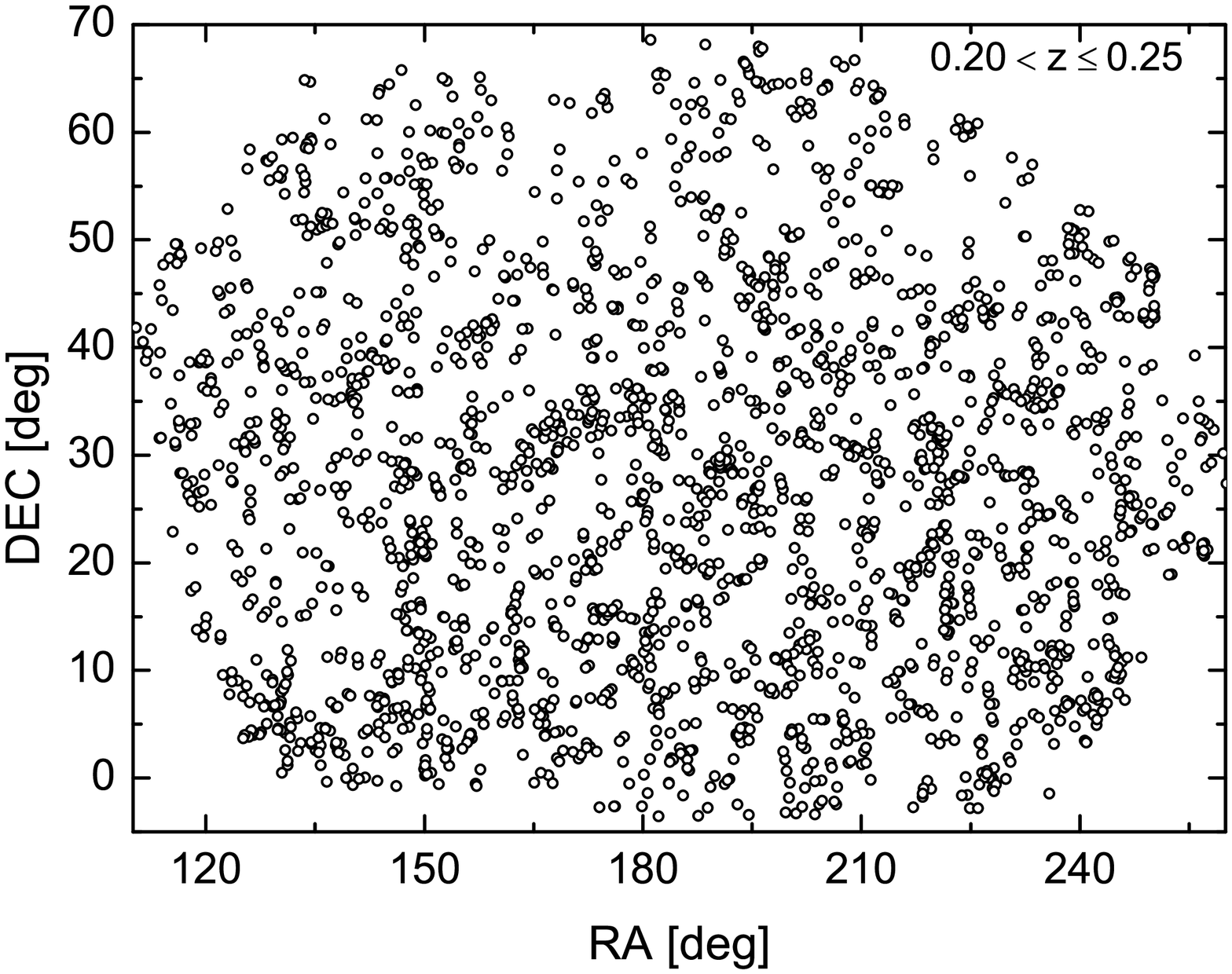} \\
\includegraphics[width=0.32\textwidth]{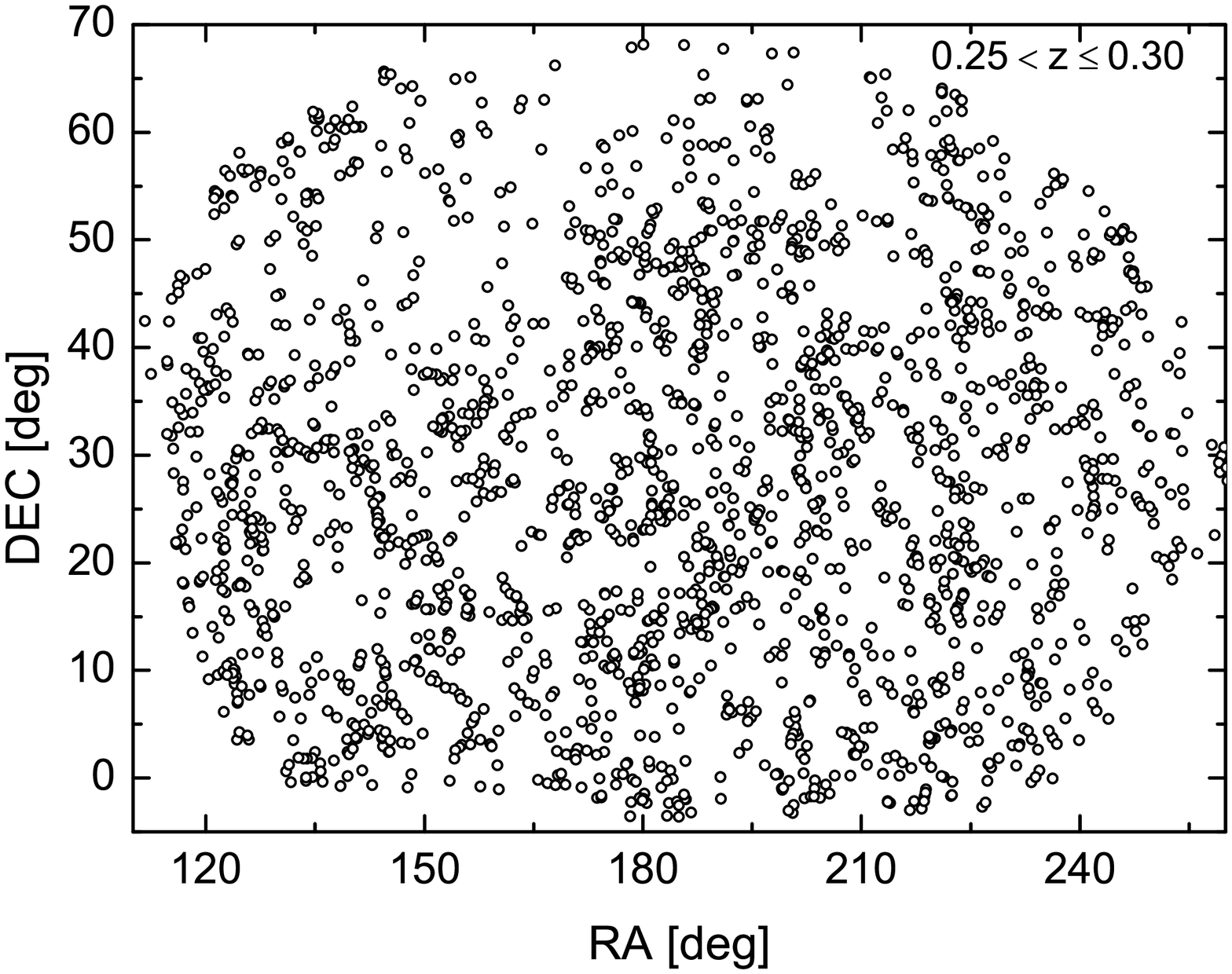}
\includegraphics[width=0.32\textwidth]{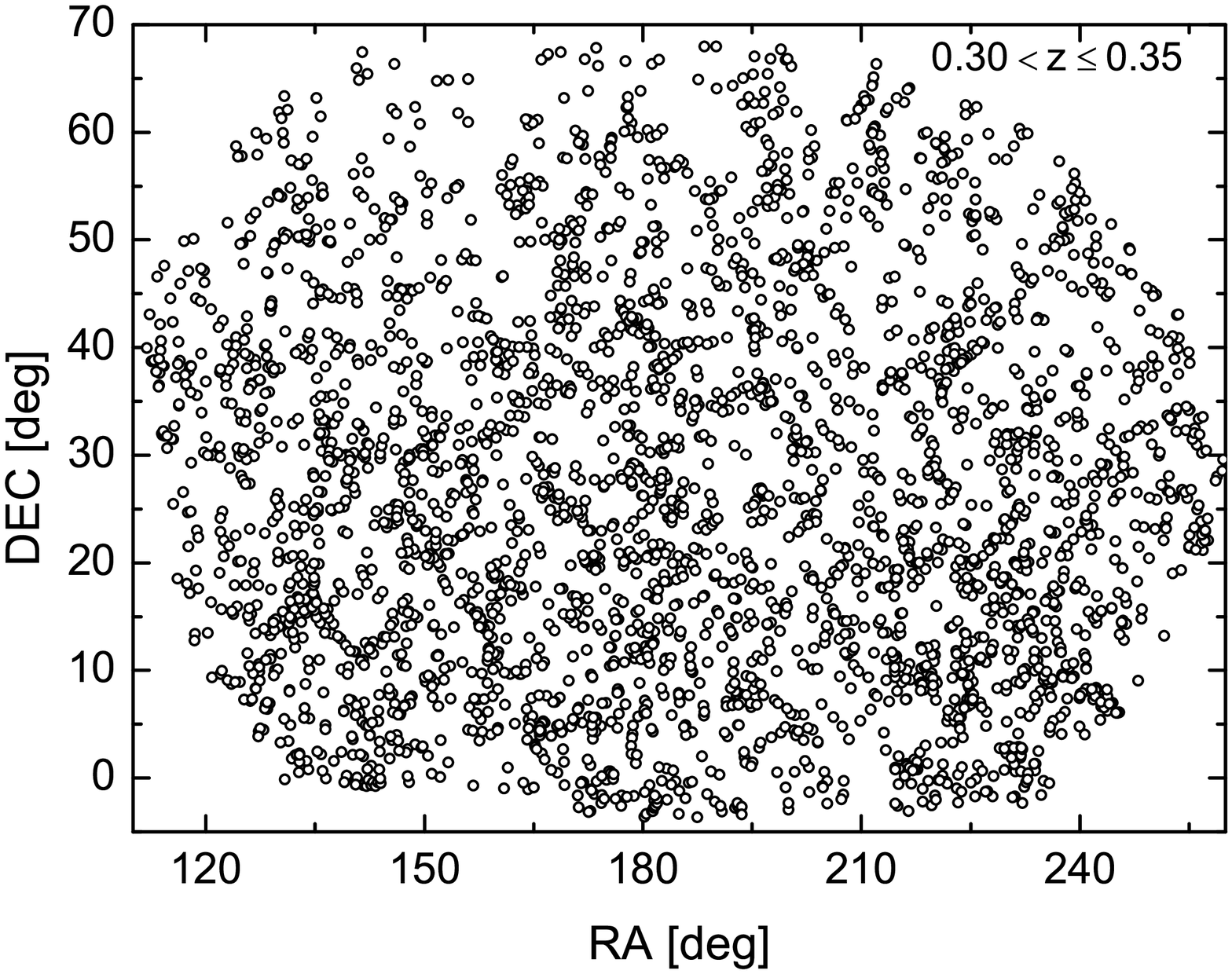}
\includegraphics[width=0.32\textwidth]{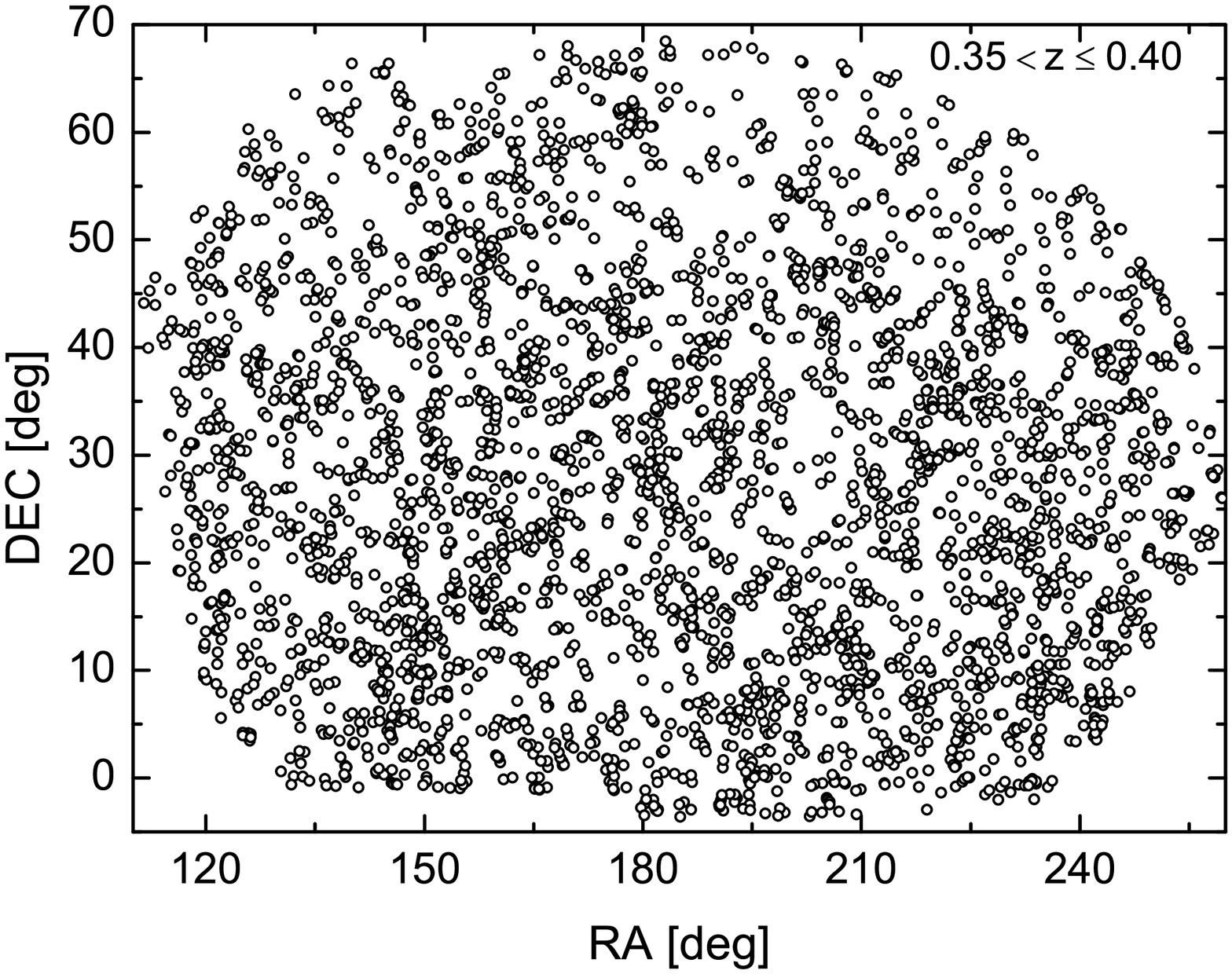}
\caption{Sky distribution of galaxy clusters in our sample
  shown in 6 redshift bins. We discard clusters in the four separated
  stripes to eliminate the boundary effect in calculation of the
  correlation function.}
\label{skycover}
\end{figure*}

In this paper, we work on the 2-point correlation function for
clusters selected from \citet{whl09} to show the power-law clustering
on small scales and the BAO feature on large scales. We describe our
cluster sample in Section 2. In Section 3 we present the correlation
functions. The power-law clustering of clusters is studied for
different richnesses of clusters. In Section 4, we analyze the BAO
feature and its cosmological implication. We verify our BAO detection
using the correlation function of the GMBCG clusters cataloged by
\citet{hmk+10}.  Conclusions are presented in Section 5.

Throughout this paper, we adopt a flat $\Lambda$CDM cosmology, with
$h=0.71$, $\Omega_{m}=0.27$, $\Omega_{\Lambda}=0.73$,
$\sigma_{8}=0.8$, where $h\equiv H_{0}/100{\rm kms^{-1}Mpc^{-1}}$.

\section{Galaxy Cluster Sample}

Using the photometric data of the SDSS DR6, \citet{whl09} identified
39\,668 galaxy clusters in the redshift of $0.05 < z < 0.60$.  This is
the largest cluster sample before the GMBCG clusters were cataloged by
\citet{hmk+10}. All clusters in \citet{whl09} contain more than eight
luminous ($M_{r}\leq-21$) member galaxies. Due to lack of
spectroscopic redshifts for distant and faint galaxies, photometric
redshifts of galaxies \citep{cbc+03} were used in the cluster-finding
algorithm. The median photometric redshift of member galaxies is
adopted to be the redshift of a cluster. By comparing the estimated
redshifts of 13\,620 clusters with the spectroscopic redshifts of the
brightest cluster galaxies (BCGs), \citet{whl09} found that the
distribution of cluster redshifts statistically have an offset of less
than 0.002 or 0.003 from the true value and the standard deviation
around 0.02. Their Monte Carlo simulations show that the
false detection rate is about 5\%. Massive clusters
of $M_{200}>2\times10^{14}{\rm M_{\odot}}$ have a high completeness 
of more than 90\% up to $z=0.42$. 

Although photometric redshifts of galaxies can be used to identify
clusters, the redshift uncertainties of clusters can smear out the
radial distribution in the large-scale structures. It has been shown
that the large uncertainty of $\sigma_{z} \sim 0.01$ is obviously the
obstacle for detection of BAO signature
\citep{bb05,zwpt08}. \citet{esf09} used the 13\,823 maxBCG clusters
with redshift uncertainty of $\sigma_{z} \sim 0.015$ and found only weak evidence
($1.4\sigma-1.7\sigma$) for the BAO peak from the 3D correlation
function.

\begin{figure}
\centering
\includegraphics[width=0.9\columnwidth]{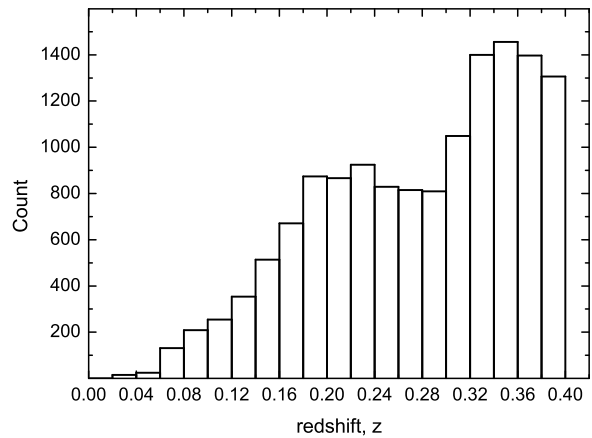}
\caption{Redshift distribution of 13\,904 clusters in our sample.}
\label{redshift}
\end{figure}

We calculate the correlation function of the selected clusters from \citet{whl09}. 
A cluster is selected only if the spectroscopic redshift of at least
one of member galaxies is available from the SDSS DR8. To
eliminate selection effect of the boundary on the correlation
function, we discard the clusters in the four small separated stripes
in the SDSS sky. Because the clusters of $z \leq 0.4$
  construct a relative complete sample \citep{whl09}, we discard all
  clusters of $z > 0.4$. There are 19\,828 clusters of $z\leq0.40$ in
total. Among them, 11\,103 clusters have their BCG
spectroscopically observed, so that we adopt the redshift of the BCG
as the cluster redshift; 2\,801 clusters have at least one of
the member galaxies (not BCG) spectroscopically observed, for which we
adopt the redshift of the galaxy or the mean of spectroscopic galaxy
redshifts as the cluster redshift. These 13\,904 clusters in the sky
area of $\sim 7\,100$ square degree shown in Figure~\ref{skycover} are
used for calculation of the correlation fuction. These are the
two-third of all clusters randomly observed for spectroscopical
redshifts. The redshift distribution of the cluster sample is shown in
Figure~\ref{redshift}. The mean redshift of the whole sample is $ z_m
= 0.276$.

\section{Correlation function of clusters}\label{sec_Method}

Two kinds of statistics are often used to describe the large-scale
clustering and BAO features. One is the 2-point correlation function
which describes the probability to find another object within a given
radius \citep{pee80}. The other is the power spectrum of object
distribution, which is the Fourier transform of the correlation
function.  Both depict the same statistical distribution properties
but in different forms. In this paper, we calculate the 2-point
correlation function $\xi(r)$ for galaxy clusters using the
Landy-Szalay estimator \citep{ls93},
\begin{equation}\label{eq.L-S}
\xi(r)=
\left[
DD(r)\frac{N_{RR}}{N_{DD}}
-2\;DR(r)\frac{N_{RR}}{N_{DR}}
+RR(r)
\right]
/RR(r),
\end{equation}
which compares the cluster pair counts with the simulated random
points. Here $DD(r)$ stands for the number of cluster pairs,
i.e. data-data pairs, within a separation annulus of $r\pm\Delta r/2$,
$DR(r)$ is the number of data-random pairs, and $RR(r)$ is the number
of random-random pairs. $N_{DD}$, $N_{DR}$ and $N_{RR}$ are the
normalization factors for the three pair counts. To minimize the
calculation noise from the random data, we simulate the random sample
16 times larger than the data sample, which covers the same sky area
(Figure~\ref{skycover}) and has the same redshift distribution
(Figure~\ref{redshift}).

We use the jackknife method to estimate the error
  covariance for the correlation function. 
  We
  divide the sky area into 32 disjoint sky sub-areas, each with
  approximately the same area as the others. We define a subsample of
  clusters by removing the clusters in only one sub-area. Correlation
  functions are calculated 32 times for 32 different subsamples. 
  The uncertainties of $\xi(r)$ at different $r$ are
  interdependent. The
  covariance matrix is then constructed as follows:
\begin{equation}
C_{ij}=\frac{N-1}{N}\sum_{k=1}^{N}\left(\xi^{k}_{i}-\overline{\xi}_{i}\right)\left(\xi^{k}_{j}-\overline{\xi}_{j}\right),
\label{err}
\end{equation}
where $N=32$ is the number of subsample, $\xi^{k}_{i}$
  represents the correlation function value of the $k^{th}$ subsample
  at the $i^{th}$ bin of $r$ values, and $\overline{\xi}_{i}$ is the
  mean value of the all 32 subsamples at the $i^{th}$ bin. The error
  bars of $\xi(r)$ are given by the diagonal elements as
  $\sigma_{i}=\sqrt{C_{ii}}$.

\begin{figure}
\centering \includegraphics[width=0.95\columnwidth]{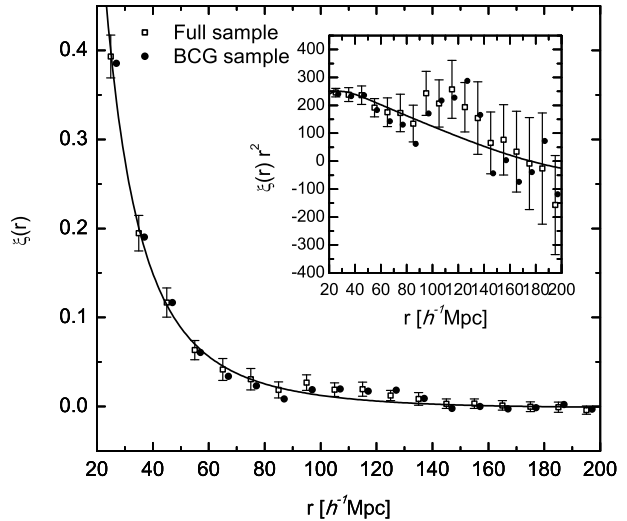}
\caption{Correlation function of 13\,904 clusters (squares)
    and 11\,103 clusters with known redshifts of BCGs (dots, shifted
    to right by $2\hMpc$ for plotting clarity). The
    solid line is the best-fit $\Lambda$CDM model without acoustic
    feature for the full sample. To show the large-scale BAO feature
    more clearly, we plot $ \xi (r) r^{2}$ in the inset.}
\label{CF}
\end{figure}

Using the 3-D spatial distribution of 13\,904 clusters with
  spectroscopic redshifts, we calculate the correlation function $\xi
  (r)$ and the uncertainty in 18 bins from $20\hMpc$ to $200\hMpc$
  (see Figure~\ref{CF}).  The correlation function on small scales
  follows a power law, and the BAO feature appears at $r\sim110
  \hMpc$. For
  comparison, we calculate the correlation function for 11\,103
  clusters which have the redshift of the BCG spectroscopically
  observed. As shown in Figure~\ref{CF}, the correlation function of
  the BCG sample is consistent with the full sample of 13\,904
  clusters.

\begin{figure}
\centering
\includegraphics[width=0.9\columnwidth,height=0.6\columnwidth]{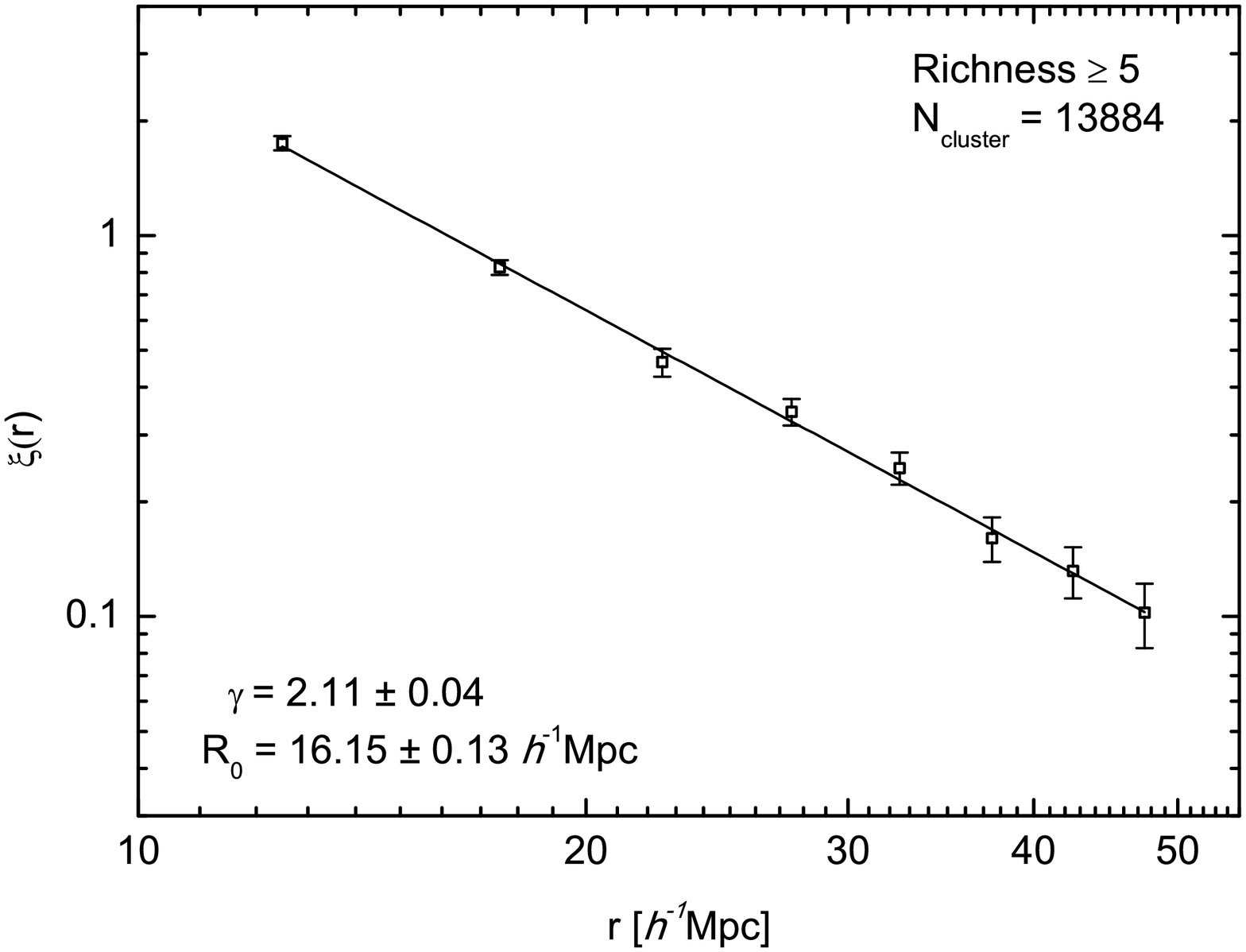}
\includegraphics[width=0.9\columnwidth,height=0.6\columnwidth]{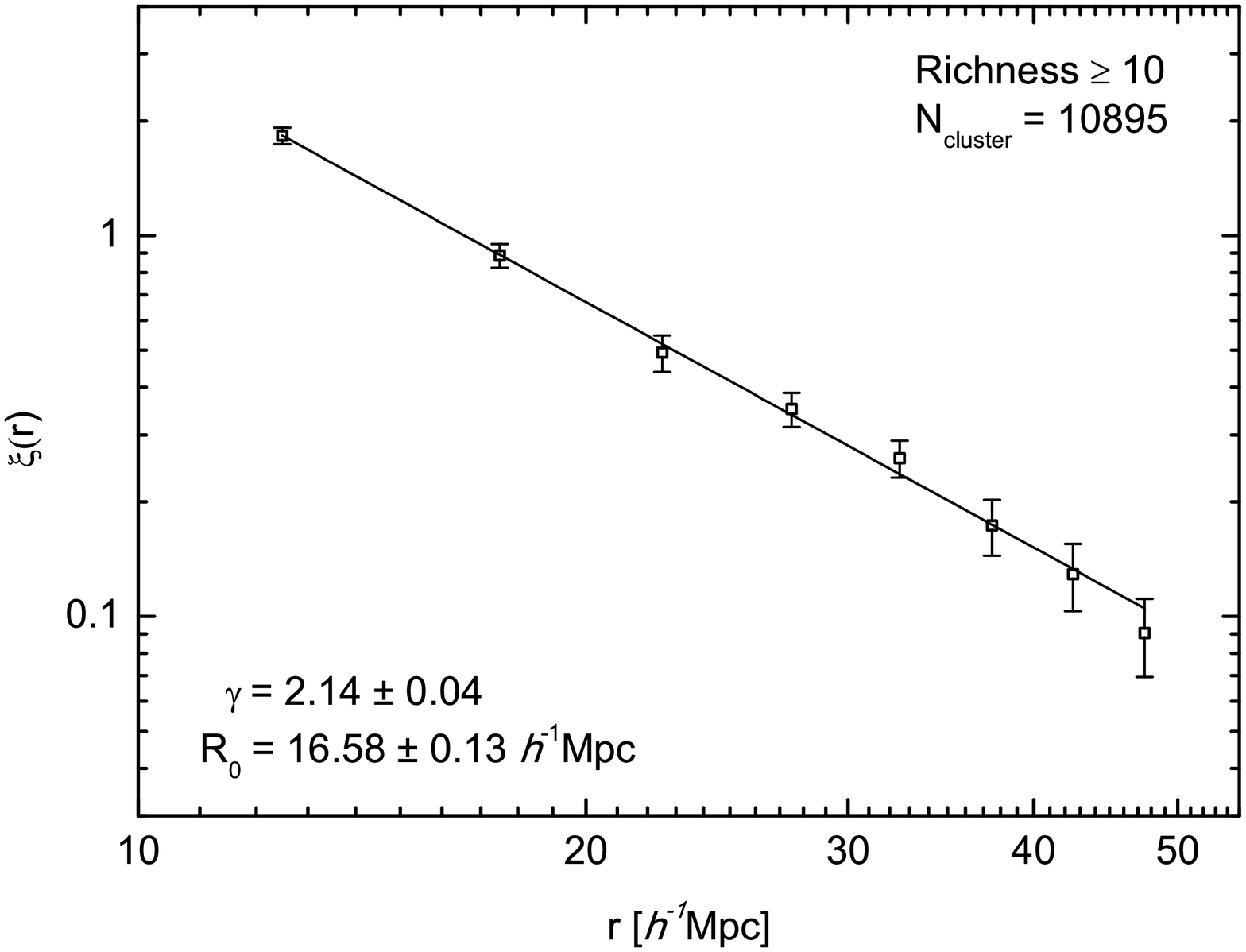}
\includegraphics[width=0.9\columnwidth,height=0.6\columnwidth]{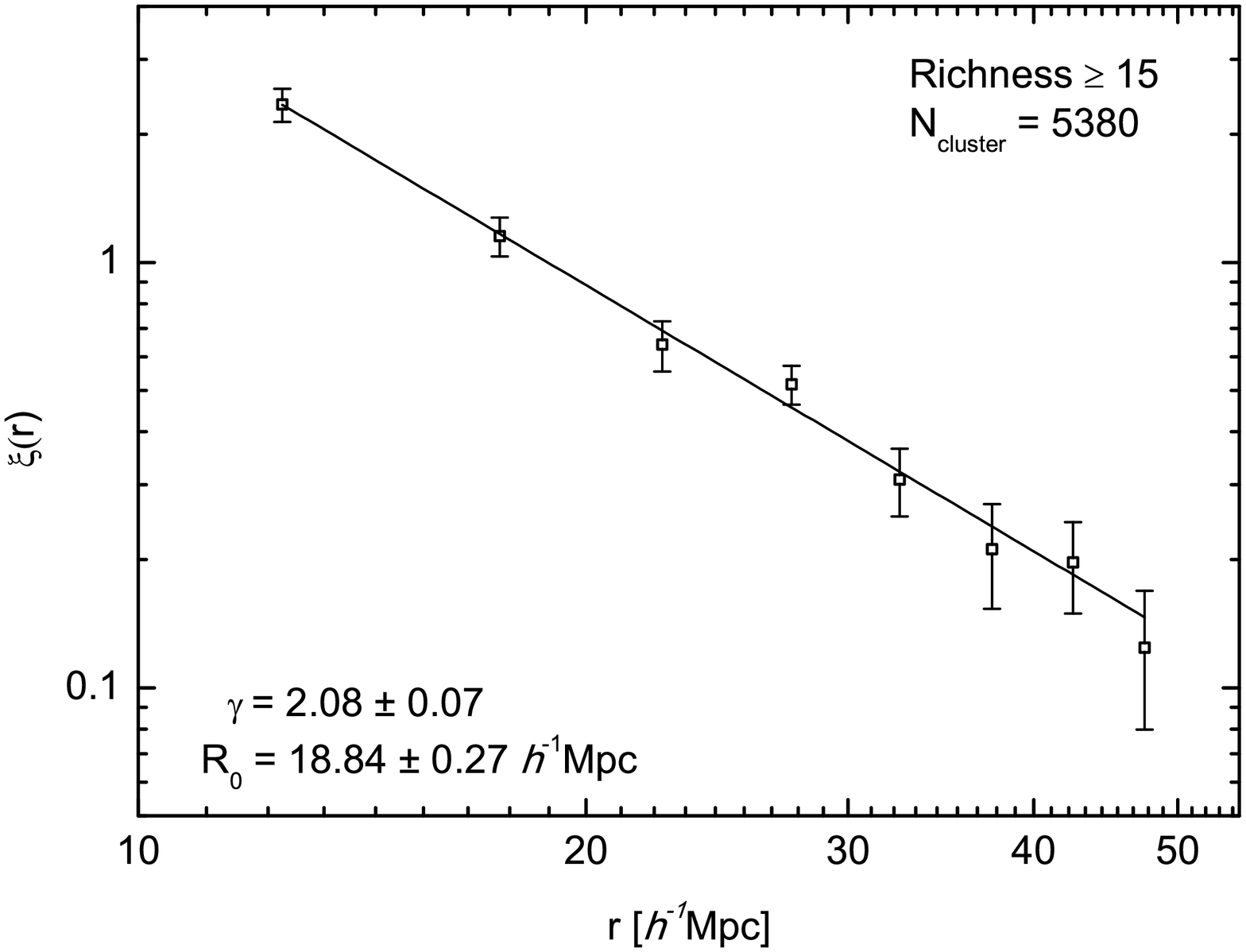}
\caption{Correlation function in the
  range $ 10\hMpc \leq r \leq 50\hMpc $ for clusters of three richness cutoffs. 
  The solid line is the best fit of a power law. }
\label{small_fit}
\end{figure}

We first analyze the correlation function on small scales.
 \citet{bs83} found that the correlation function of the Abell
  clusters is consistent with a power law on scales of less than
  several tens of $\hMpc$. They also showed the correlation function
  is richness dependent, i.e. the correlation length increases with
  cluster richness. Their conclusions were confirmed by others using
  various cluster samples \citep{bdh+03,cde+97,cgb+00,lp02,esf09}.
  The correlation function of the small scales ($r \leq 50\hMpc$) can
  be fitted by the power law,
\begin{equation}\label{eq3.1}
\xi(r)=\left( \frac{r}{R_{0}}\right)^{-\gamma},
\end{equation}
where $R_{0}$ is the correlation length, $\gamma$ is the power law
index.  We calculate the correlation functions for our cluster sample
with different thresholds of cluster richness (see
Figure~\ref{small_fit}). Fitting the data with the power law in the
range of $ 10\hMpc \leq r \leq 50\hMpc$, we get the values of $R_{0}$
and $\gamma$, as $R_{0}=18.84\pm0.27 \hMpc$ and $ \gamma =
2.08\pm0.07$ for 5\,380 clusters with a richness of $R\geq15$;
$R_{0}=16.58\pm0.13 \hMpc$ and $ \gamma = 2.14\pm0.04$ for 10\,895
clusters with $R\geq10$; and $R_{0}=16.15\pm0.13\hMpc$ and $ \gamma =
2.11\pm0.04$ for 13\,884 clusters with $R\geq5$. Our results are
consistent with but more accurate than any previous works
\citep[i.e.][using maxBCG clusters]{esf09}. We find that the power
index $\gamma$ almost does not change with the cluster richness, while
the correlation length changes from $18.84 \hMpc$ for very rich
clusters to $16.15 \hMpc$ for less rich clusters. We also find that
the correlation length $R_{0}$ of clusters is considerably larger than
those obtained from the galaxy samples, e.g.  $5.91\hMpc $ by
\citet{zwz+04} using the volume limited SDSS galaxy sample, $
R_{0}=5.05\hMpc $ by \citet{hmc+03} using the 2dF galaxies.

\section{Baryon acoustic oscillations and cosmological modeling}
\label{method}
The correlation function shown in Figure ~\ref{CF} has an obvious
excess around the $ r\sim 110 \hMpc$ over the non-baryon model, which
is the BAO feature in the large-scale structure of the
universe. 
Here, we analyse the
BAO feature using a $\Lambda$CDM model.
%
To get the theoretical curve, we first obtain the linear matter power
spectrum. The damped BAO feature has a non-linear evolution to each
redshift in a $\Lambda$CDM model. We compute the redshift-space
correlation functions at the central redshift of each counting bin
between $z=0-0.4$ (see Figure~\ref{redshift}), and then weight these
functions using the corresponding number counts, i.e., the sample
redshift distribution $n(z)$, to obtain the ``averaged'' correlation
function. We finally fit this theoretical BAO model to real
measurements, and constrain the cosmological parameters.

First, we compute the linear matter power spectra at each redshift
using {\sc cmbfast} \citep{zs00}. The damping of the BAO feature due
to nonlinear evolution is approximated by an elliptical Gaussian
\citep{esw07}
\begin{equation} 
\tilde{P}(\mbox{\textbf{\emph{k}}}) = 
\left(P_\mathrm{lin}(k)-P_\mathrm{nw}(k)\right)
\exp{\left (-\frac{k_{\perp}^2\Sigma_{\perp}^2}{2}-
  \frac{k_{\parallel}^2\Sigma_{\parallel}^2}{2}\right )},
\label{eq.pk}
\end{equation}
where $P_\mathrm{lin}(k)$ is the linear matter power spectrum, and
$P_\mathrm{nw}(k)$ is the no-wiggle approximation of the linear matter
power spectrum \citep{eh98}. The rms of radial displacements across
the line of sight is given by $\Sigma_{\perp}=\Sigma_0 G$, where
$\Sigma_0 = 11\hMpc$, which is scaled from $12.4\hMpc$ \citep{esw07}
at $\sigma_8=0.9$ to $\sigma_8=0.8$ \citep{ldh+10}, and $G$ is the
linear growth factor normalized as $G(z=0)=0.76$ in the reference
cosmological model. The rms of radial displacements along the line of
sight is $\Sigma_{\parallel}=\Sigma_0 G(1+f)$), where $f=d(\ln
G)/d(\ln a)\sim \Omega^{0.6}$ is the growth rate.  Even though
Equation~(\ref{eq.pk}) only involves the linear power spectra, the
damping of the oscillating feature
$\tilde{P}(\mbox{\textbf{\emph{k}}})$ is calculated based on $N$-body
simulations. Hence, $\tilde{P}(\mbox{\textbf{\emph{k}}})$ needs not be
subject to nonlinear mapping of the power spectrum again, though the
difference is small in practice, whether it is mapped or not.

The nonlinear matter power spectrum is then given by the sum of the
nonlinear no-wiggle matter power spectrum $P_\mathrm{nl,nw}(k)$ and
the damped BAO feature $\tilde{P}(\mbox{\textbf{\emph{k}}})$
\begin{equation} 
\label{eq.halofit}
P_\mathrm{nl}(\mbox{\textbf{\emph{k}}})=
\tilde{P}(\mbox{\textbf{\emph{k}}}) + P_\mathrm{nl,nw}(k),
\end{equation}
where we choose to use the \citet{pd96} fitting formula to calculate
$P_\mathrm{nl,nw}(k)$ for simplicity. Because the scales of interest
are fairly large, different nonlinear fitting formulae produce very
similar correlation functions at the end.

We next inlcude the linear redshift distortion effect \citep{kai87}
and nonlinear redshift distortion effect, along with the bias of
galaxy cluster $b$, to obtain the power spetrum of galaxy clusters,
$P_\mathrm{c}(k,\mu)$, in the redshift space
\begin{equation} 
\label{eq.pkg}
P_\mathrm{c}(k,\mu) = P_\mathrm{nl}(k,\mu) 
\frac{b^2 (1+\beta\mu^2)^2}{1+k^2\mu^2\sigma_p^2/2H^2},
\end{equation}
where $\mu$ is the direction cosine.  $\beta=f/b$ is the linear
redshift distortion factor, $\sigma_p = 400 \mbox{km\,s}^{-1}$ denotes
the pairwise velocity dispersion \citep[see, e.g.,][]{pcn+01}.  The
denominator of Equation (\ref{eq.pkg}) models the finger-of-God effect
or the nonlinear redshift distortion, which suppresses the power blow
the cluster scale. Even though we measure correlations between
clusters, the results may still be affected by the nonlinear redshift
distortion because the redshift of each cluster is represented by the
redshift of a single galaxy in most cases. In practice, however, we
find that the nonlinear redshift distortion term in Equation
(\ref{eq.pkg}) is unimportant for this work.

\begin{figure}
\centering
\includegraphics[width=0.9\columnwidth]{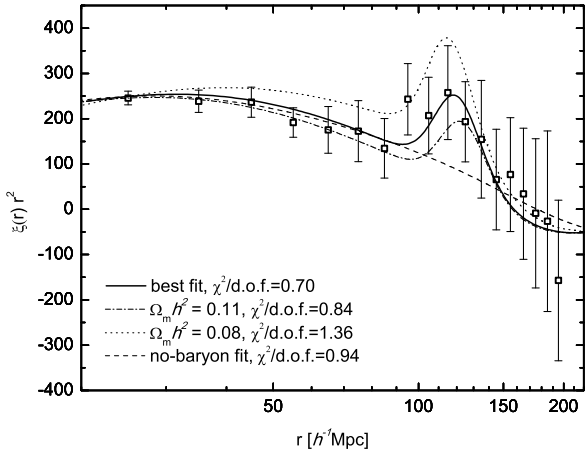}
\caption{The observed correlation function of 13\,904 clusters
    in 18 bins from $20\hMpc $ to $ 200\hMpc$, together with the
    best-fit model curves with the BAO features (solid line) and
    without the BAO features (dashed line). The dot-dashed line
    ($\Omega_{m}h^{2} = 0.11$) and dot line ($\Omega_{m}h^{2} = 0.08$)
    outline two other models.}
\label{bestfit}
\end{figure}

We multiply $b^2$ into the subsequent parentheses in Equation
(\ref{eq.pkg}). By Fourier transforming this equation, we obtain the
theoretical redshift-space two-point correlation function (monopole):
\begin{equation} 
\label{eq.cf}
\xi=b^2 \xi_1+b \xi_2 + \xi_3,       
\end{equation}
which consists of the components of the correlation function that have
different dependence on the {\bf cluster} bias. In this way, we can
vary $b$ in the data fitting process efficiently.

Fitting the observed correlation function with theoretical curves is
to compute $\chi^{2}$ using the full covariance matrix for a grid of
parameters $\Omega_m h^2, D_v(0.276)$ and $b$, where $D_v(z)$ is the
reduced distance at the mean redshift of our sample
$z_m=0.276$. $D_v(z)$ is firstly defined in \citet{ezh+05} as,
\begin{equation}\label{eq:D}
D_v(z) = \left[ D_A(z)^2 {cz\over H(z)}\right]^{1/3},
\end{equation}
\noindent where $H(z)$ is the Hubble parameter and $D_A(z)$ is the
comoving angular diameter distance. For the reference cosmology,
$\Omega_m=0.27$, $\Omega_{\Lambda}=0.73$, $h=0.71$ and
$D_v^{ref}(0.276)=1072.5 {~\rm Mpc}$. Note, the amplitude is not an
independent parameter but its effect is included in the parameter $b$
here. The other involved cosmological parameters have been fixed to
the WMAP7 \citep{ldh+10} best-fit values: $\Omega_b=0.0449$,
$n_s=0.96$ and $\sigma_8=0.8$.

\begin{figure}
\centering
\includegraphics[width=0.95\columnwidth]{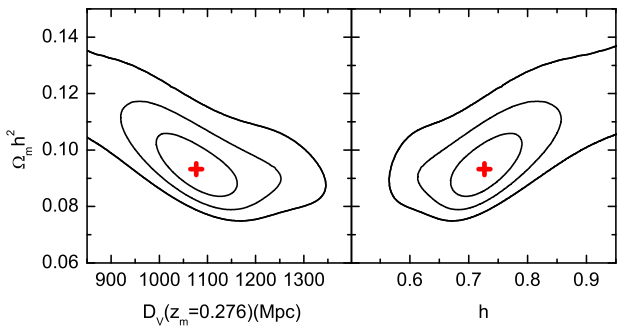}
\caption{Likelihood contours for the best-fit as $\Omega_mh^2=0.093$
  and $D_v=1077 {\rm Mpc}$ as a function of $D_{v}(z_{m}=0.276)$ and
  $\Omega_{m}h^{2}$ {\it (left)}, and the likelihood contours for
  $h$ and $\Omega_{m}h^{2}$ {\it (right)}. From the inner to the
  outer, contours corresponding to $1 \sigma$, $2 \sigma$, and $3
  \sigma$ respectively.}
\label{likely}
\end{figure} 

We do the $\chi^2$ search by scanning over a large table of $\Omega_m
h^2$, $D_v(0.276)$ and b($0.05 \leq \Omega_{m}h^{2} \leq 0.15$; $0.60
\leq s=D_v^{ref}(0.276)/D_v(0.276)\leq 1.30$; $0.01\leq b \leq
3.75$). We then marginalize over $b$ by integrating the probability
distribution $P(\Omega_m h^2,D_v,b) \propto Exp[-0.5\chi^2]$ along $b$-axis to
give constraints on $\Omega_m h^2$ and $D_v$.  The best-fit curve is
shown in Figure~\ref{bestfit}, with the best-fit of $\chi^2=10.52$ on
$15$ degrees of freedom (18 data points and 3 parameters; the reduced
$\chi^2=0.70$). The best-fit pure CDM model without the BAO feature is
also presented (the dashed line), which has a $\chi^2=14.02$ and is
rejected at $1.87\sigma$.  The marginalized constraints on $\Omega_m
h^2$ and $D_v(0.276)$ are $D_v=1077\pm55(1\sigma) {~\rm Mpc}$ and
$\Omega_mh^2=0.093 \pm0.0077(1\sigma)$ respectively, as shown in
Figure \ref{likely}.  Assuming a flat cosmology with a cosmological
constant model, there are only 2 degree of freedom of parameters in
$D_v$ (indeed in $H(z)$). The constraints can be directly imposed on,
say, $\Omega_mh^2-h$, so that we get $h=0.73\pm0.039(1\sigma)$ (see
Figure \ref{likely}).

\begin{figure}
\centering
\includegraphics[width=0.9\columnwidth]{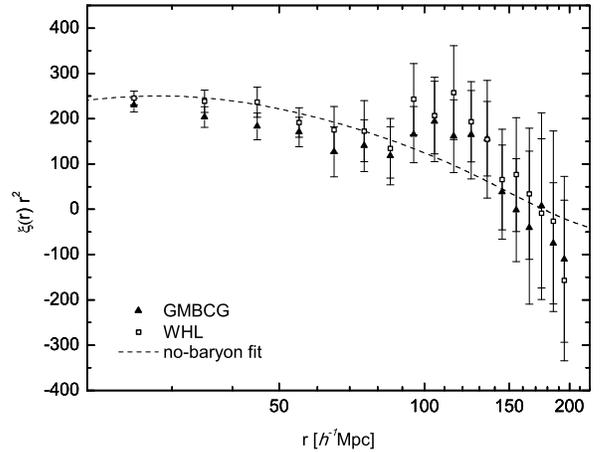}
\caption{The correlation function of clusters from the GMBCG sample
  \citep{hmk+10} compared with that from \citet{whl09} cluster sample.
  The BAO features from the two samples are consistent.}
\label{cf_GMBCG}
\end{figure} 

\citet{bkb+11} detected the BAO peak in the WiggleZ correlation functions 
for redshift slices of width $\Delta z =0.4$, which is the same width with our cluster sample. 
\citet{bkb+11} fit these correlation functions using a four-parameter model, the 
fitting results are in agreement with WMAP 7-year value on $1\sigma$ level \citep[see Table 2 in][]{bkb+11}. Our cluster correlation function shows
  a more enhanced BAO peak than expected from the CMB model (see
  Figure~\ref{bestfit}). 
From the model-fitting to the correlation function of galaxy clusters, 
we get constraints of the distance-scale measurements 
$D_v(0.276)$ and $h$ which are 
  consistent with WMAP 7-year results, while the constraint on 
  $\Omega_{m} h^{2}$ is different from the widely
  accepted value obtained from the WMAP 7-year data \citep[$\Omega_{m}
    h^{2}=0.133$,][]{ldh+10}.  We have tested our analyzing algorithm by
  calculating and fitting the correlation functions of 160 LasDamas SDSS
  LRG mock catalogs (McBride et al. 2011, in preparation), and we found that  
  our analysis method does not introduce any bias in the parameter 
  estimation (see the Appendix for details).

In the very late stage of this work, \citet{hmk+10} published a new
cluster catalog based on the BCG recognization and red-sequence
features of galaxies, in which they identified 55\,424 rich clusters
in the redshift range $0.1 < z < 0.55$ from the SDSS DR7 data. Using
their clusters, we get a sample of 15\,074 clusters with available
spectroscopical redshifts, which has almost the same sky coverage and
redshift range as our sample. The correlation function of these
15\,074 clusters as shown in Figure~\ref{cf_GMBCG} is in good
agreement with that from our cluster sample selected from \citet{whl09}.

\section{Summary}

Galaxy clusters identified by \citet{whl09} from the SDSS cover a
large scale area and have redshifts up to $z\sim0.6$. We calculate the
2-point correlation function of a sample of clusters selected from
\citet{whl09} to study the large-scale structure of the universe. To
avoid the smearing effect from photometric redshift errors, clusters
with at least one member galaxy spectroscopical observed for redshifts
are selected.  Our cluster sample contains 13\,904 clusters of
$z\leq0.4$.  The correlation function on the scales $ 10\hMpc \leq r
\leq 50\hMpc $ of this cluster sample follows a power law. The
power-law index about $\gamma=2.1$ is almost the same for the
richer clusters and poorer clusters. The richer clusters have a larger
correlation length. We get $R_0= 18.84\pm0.27\hMpc$ for clusters of
richness $R\geq15$ and $R_0= 16.15\pm0.13\hMpc$ for clusters of
$R\geq5$.  This is consistent with but more accurate than previous results.

The correlation function on large-scale of $ 20\hMpc \leq r \leq
200\hMpc $ shows the baryon acoustic peak around $r \sim
110\hMpc$ with a significance of $1.9\sigma$. We fit
the observed correlation function with a parametrized theoretical
curve, which is determined by the physical matter density parameter
$\Omega_{m} h^{2}$, the stretch factor $s$, and the galaxy bias.  We
obtain the constraints $\Omega_{m} h^{2}=0.093\pm0.0077(1\sigma)$, 
$D_v(0.276)=1077\pm55(1\sigma) {~\rm Mpc}$ and
$h=0.73\pm0.039(1\sigma)$.

Because the detected BAO peak in our cluster sample is 
stronger than expected, the estimated matter 
density parameter is more than $3\sigma$ lower than the 
WMAP 7-year result. It is unclear why the cluster correlation function shows the stronger 
BAO signal. Nevertheless, our results demonstrate that one 
can detect the BAO peak in the cluster correlation function at the expected scale.
  Future larger all sky spectroscopic galaxy and cluster
  surveys, such as BigBOSS \citep{saa+11}, will provide deeper and
  more uniform spectroscopic samples of clusters for large-scale structure
  analysis, and can be used to verify the amplitude of BAO peak we
  detected.


\begin{acknowledgments}
We thank Y. Y. Zhou, F. S. Liu, F. Beutler, Q. Wang and X. Y. Gao for useful
comments.  The authors are supported by the National Natural Science
foundation of China (10821061, 10833003 and 11033005) and the National
Key Basic Research Science Foundation of China (2007CB815403,
2010CB833000).  H. Zhan and L. Sun are supported by the Bairen program
from the Chinese Academy of Sciences.
%
%
\end{acknowledgments}

\appendix
\section{Testing the analyzing algorithm}
We calculate and fit the correlation functions of 160 LasDamas mock
catalogs to test our cosmological analyzing algorithm.  The LasDamas
simulation uses a single
cosmological model with the same WMAP 5-year values of $\Omega_m=0.25$, $\Omega_\Lambda=0.75$,
$\Omega_b=0.04$, $h_0=0.7$, $\sigma_8=0.8$ and $n_s=1.0$ \citep{kdn+09}. To build realistic
SDSS mock catalogs, the LasDamas team places galaxies inside dark
matter halos using a Halo Occupation Distribution with parameters from
observed SDSS catalogs.

\begin{figure}
\centering
\includegraphics[width=0.9\columnwidth]{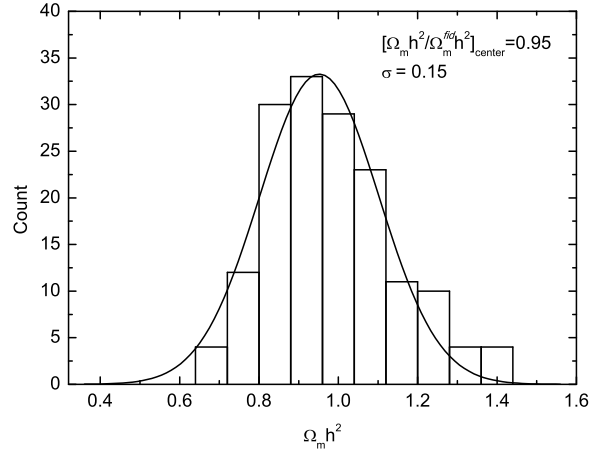}
\caption{The distribution of the ratio of the fitting parameter
  $\Omega_m$ and the fiducial $\Omega_{m}^{fid}$. The solid line is
  the best fit of a Gaussian function, the Gaussian fitting center is
  $[ \Omega_m h^2/\Omega_{m}^{fid} h^2 ]_{center}=0.95$, with a $\sigma=0.15$. 
  The mean value of 160 mock catalog fitting results is 
  $[ \Omega_m h^2/\Omega_{m}^{fid} h^2 ]_{mean}=0.99$.}
\label{las_omega}
\end{figure} 
\begin{figure}
\centering
\includegraphics[width=0.9\columnwidth]{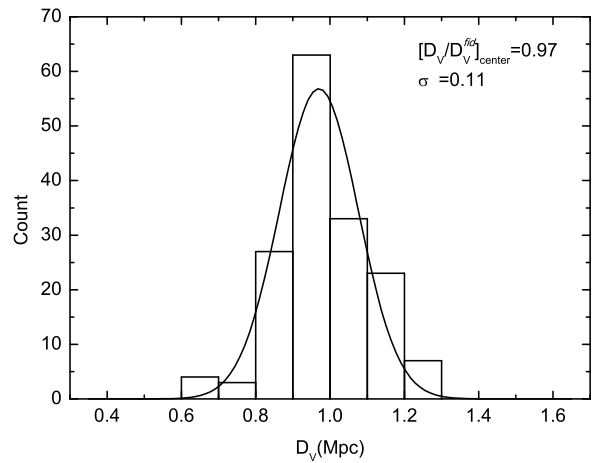}
\caption{The distribution of the ratio of the fitting parameter $D_v$
  and the fiducial $D_{v}^{fid}$. The solid line is the best fit of a
  Gaussian function, the Gaussian fitting center is
  $[ D_v/D_{v}^{fid} ]_{center}=0.97$, with a $\sigma=0.11$.
  The mean value of 160 mock catalog fitting results is 
  $[ D_v/D_{v}^{fid} ]_{mean}=0.98$.}
\label{las_scale}
\end{figure} 
\begin{figure}
\centering
\includegraphics[width=0.9\columnwidth]{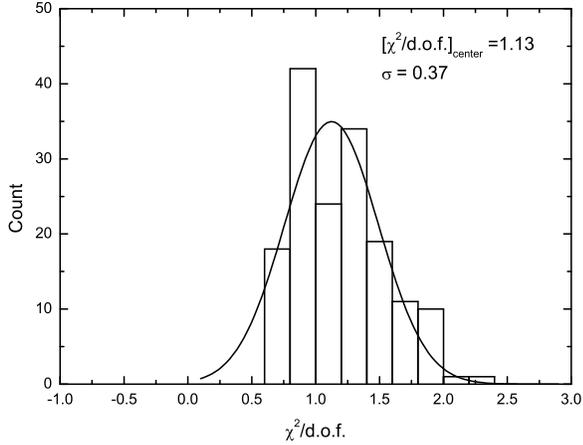}
\caption{The distribution of reduced $\chi^2$. The solid line is the
  best fit of a Gaussian function, the Gaussian fitting center is
  $\chi^2=1.13$, with a $\sigma=0.37$.}
\label{las_chi}
\end{figure} 

We build 160 testing mock catalogs by selecting 12\,488 galaxies
randomly from 160 different LasDamas LRG realizations which contain
about 31\,500 LRG galaxies ($M_g < -21.8$) respectively. All testing
mock catalogs have the same footprint and redshift distribution with
our cluster data catalog in the redshift region $0.16 \leq z \leq
0.40$. The theoretical correlation function curve is calculated for
the 160 testing mock catalogs with the same fiducial cosmological
parameters of the LasDamas simulation using the method mentioned in
Section \ref{method}.

Correlation functions are calculated for the 160 testing LRG mock
catalogs respectively.  By fitting these correlation functions with
the theoretical curve, we find our analyzing algorithm works well. The
statistical distribution of fitting parameters $\Omega_m$ and $D_v$
are shown in Figure \ref{las_omega} and Figure \ref{las_scale}. We fit
these distributions with a Gaussian function, the fitting centers are
$\Omega_m/\Omega_{m}^{fid}=0.95\pm0.15$ and
$D_v/D_{v}^{fid}=0.97\pm0.11$, which is consistent with the fiducial
cosmology of the LasDamas mock catalogs. The distribution of 160
reduced $\chi^2$ is also shown in Figute \ref{las_chi}.

\bibliographystyle{apj}
\bibliography{bibfile}
\end{document}